\documentclass[12pt]{iopart}

\usepackage{graphics}
\usepackage{graphicx}
\usepackage{iopams} 
\usepackage{siunitx}
\usepackage{cite}
\begin{document}
\newcommand{\kpt}{\mathbf{k}}
\newcommand{\tiKS}{|\mu_{\mathbf{k}n}\rangle}
\newcommand{\tdKS}{|v_{\mathbf{k}n}\rangle}
\newcommand{\tdFL}{|\phi_{\kpt n}\rangle}
\newcommand{\tiFL}{|\varphi_{\kpt n}^\eta\rangle}
\newcommand{\dFKS}{\widetilde{d}_{\kpt ni}(\eta) }
\newcommand{\dFKSj}{\widetilde{d}_{\kpt nj}(\gamma) }
\newcommand{\kplus}{\mathbf{k}_i^\sigma}
\newcommand{\SuppInfo}{Supplemental Material}
\newcommand{\rrr}{\textcolor{red}{[ref]}}
\newcommand{\kgrid}[3]{#1$\times$#2$\times$#3}

%%%%%%%%%%%%%%%%%%%%%%%%%%%%%%%%%%%%%%%%%%

\title[Floquet approach to nonlinear optics]{Floquet formulation of the dynamical Berry-phase approach to nonlinear optics in extended systems} 

\author{Ignacio M. Alliati \& Myrta Gr\"uning}
\address{School of Mathematics and Physics, Queen’s University Belfast, Belfast BT7 1NN, Northern Ireland, United Kingdom}
\address{European Theoretical Spectroscopy Facility (ETSF)}
\ead{m.gruening@qub.ac.uk}

\begin{abstract}
We present a Floquet scheme for the ab-initio calculation of nonlinear optical properties in extended systems. This entails a reformulation of the real-time approach based on the dynamical Berry-phase polarisation [Attaccalite \& Gr\"uning, PRB 88, 1–9 (2013)] and retains the advantage of being non-perturbative in the electric field. The proposed method applies to periodically-driven Hamiltonians and makes use of this symmetry to turn a time-dependent problem into a self-consistent time-independent eigenvalue problem. We implemented this Floquet scheme at the independent particle level and compared it with the real-time approach. Our reformulation reproduces real-time-calculated 2$^{nd}$ and 3$^{rd}$ order susceptibilities for a number of bulk and two-dimensional materials, while reducing the associated computational cost by one or two orders of magnitude.
\end{abstract}

\vspace{1pc}
\noindent{\it Keywords}: Floquet theory, nonlinear optics, second harmonic generation, Berry-phase polarisation, periodically-driven quantum systems

\section{Introduction}\label{sc:intro}
Nonlinear optical spectroscopies are an invaluable tool for investigating materials properties. For example, second harmonic generation (SHG), due to its sensitivity to electric fields and symmetry properties, has been traditionally used on surfaces and interfaces~\cite{PhysRev.174.813,annurev.ms.16.080186.000441,annurev.pc.40.100189.001551} and more recently for the characterisation and imaging of two-dimensional flakes~\cite{doi:10.1021/nl401561r,doi:10.1126/science.1250564,doi:10.1021/nn500228r,Woodward_2016}. In addition, SHG is being used as a highly sensitive probe of magnetic ordering in atomically thin materials and in multiferroics~\cite{Sun2019b,doi:10.1021/acs.nanolett.2c00212,Chauleau2017}. Due to the sensitivity to changes in the electric polarisation, SHG can also probe the dynamic of excited systems, tracking---for instance---the formation of excitons, exciton-phonon coupling and demagnetisation of antiferromagnets~\cite{PhysRevB.61.14716,adma.201705190}. Further, nonlinear optical properties of materials are intensely investigated for applications to quantum technologies, optical frequency metrology and optoelectronics. Recent developments have highlighted the crucial role low dimensionality can play in this field \cite{Liu2017,Autere2018,You2018,Dogadov2022}, for instance, strengthening the nonlinear response as well as facilitating the integration into devices. Other advances include epsilon-near-zero media \cite{Reshef2019}, the new generation of infrared nonlinear optical materials \cite{Abudurusuli2021} and the use of structured light \cite{Buono2022}.
The development of \textit{ab-initio} approaches to nonlinear optical properties is then of utmost importance for interpreting experimental measurements and guiding the design of functional materials. While the \textit{ab-initio} prediction of optical properties in the linear regime is well-established \cite{Strinati1988,Onida2002}, the theoretical description of nonlinear processes still presents challenges concerning its complexity and the associated computational cost. 

The theoretical methods available in the literature for the calculation of nonlinear optical properties are either perturbative or non-perturbative. Perturbative approaches are often extensions of frameworks that proved successful for linear optics. For instance, Sipe \textit{et al.} presented a scheme for the calculation of nonlinear optical response of semiconductors within the independent particle approximation (IPA) and derived expressions for the second order susceptibility, $\chi^{(2)}$ \cite{Sipe1993}. A study by Dal Corso \textit{et al.} introduced a Sternheimer approach for the second order response of insulators within the local density approximation (LDA) \cite{Corso1996a}. Luppi \textit{et al.} derived perturbative expressions for $\chi^{(2)}$ in extended systems from time-dependent density functional theory (TD-DFT) \cite{Luppi2010}. The latter included excitonic effects by means of a long-range contribution to the exchange-correlation kernel and proved valid for weakly bound excitons \cite{Luppi2010}. Finally, the inclusion of many-body effects at the Bethe-Salpeter equation (BSE) level warranted a few attempts up to the second order \cite{Leitsmann2005,Chang2001}. Unfortunately, perturbative approaches require a specific formulation for each order in the response one intends to calculate and their generalisation to higher orders is not straightforward. Indeed, the resulting expressions for nonlinear susceptibilities become extremely complex with increasing orders in the perturbation and increasing levels of theory as regards correlation.

At variance, non-perturbative approaches involve explicit time propagation and can describe nonlinear phenomena to several orders in the electric field simultaneously, thus offering a convenient workaround to the shortcomings described above. Moreover, they are flexible in the sense that including many-body effects amounts to just adding the relevant operators into the effective Hamiltonian. In these methods, the integration of an equation of motion (EOM) allows for the calculation of the dynamical polarisation, from which susceptibilities to any order (in principle) can be extracted. The quantity evolved in the EOMs varies among the different time-propagation methods. For instance, TD-DFT implies the time evolution of the electron density and is typically applied to isolated systems \cite{Yabana1996,Marques2004,Casida2009,Casida2012,Laurent2013}. Propagating the Green's function was proposed in the so-called Kadanoff-Baym equations (KBE) \cite{Kadanoff}. A simplification of the KBE using the time-diagonal of said Green's function, i.e., the density matrix, was proposed by Attaccalite, Gr\"uning and Marini \cite{Attaccalite2011}. Subsequently, Attaccalite and Gr\"uning proposed a scheme based on evolving the periodic part of the Bloch functions \cite{Attaccalite2013}. Crucially, this method is valid for systems with periodic boundary conditions (PBCs) since it is based on the modern theory of polarisation \cite{Vanderbilt1993,Vanderbilt1993a,Resta1993} and uses the Berry phase formulation of the dynamical polarisation \cite{Souza2004} (see Section \ref{sc:methods}). This real-time approach \cite{Attaccalite2013} has been successfully applied for the calculation of nonlinear optical properties in extended systems \cite{Attaccalite2015,Attaccalite2017,Attaccalite2018,Attaccalite2019,Wei2019}.

The main drawback of time-propagation approaches lies in their elevated computational cost, which results from the short time steps and long simulated times required. This implies repeating a handful of operations, e.g., building a Hamiltonian matrix, for tens of thousands of time steps. The computational cost of these schemes renders the calculation of nonlinear optical properties prohibitively costly in many cases, certainly for large systems and complex materials. Therefore, finding alternative formulations and methods that could alleviate these computational demands is of utmost importance. In order to tackle this challenge, the use of time-periodicity and Floquet theory offers an interesting avenue to explore. In periodically-driven systems, one could avoid the explicit integration of the EOMs and reformulate them as a time-independent eigenvalue problem \cite{Shirley1965} by invoking Floquet's theorem. This has been attempted at the TD-DFT level \cite{Telnov1997,Chu2004} and intensely debated \cite{Maitra2002,Samal2006,Maitra2007,Kapoor2013}. However, to the best of our knowledge, it has been only applied to atomic and molecular systems \cite{Telnov1997,Chu1989,Chu2004,Salek2005}. 

In this work, we introduce an efficient Floquet approach to nonlinear optics valid for extended systems. Our scheme works under PBCs since we use the Berry-phase polarisation, and its conjunction with Floquet theory defines the originality of our contribution. We implemented our method at the IPA level and achieved a sizeable computational advantage compared to the real-time approach \cite{Attaccalite2013} while retaining its main benefits, i.e., it is non-perturbative in the electric field and offers flexibility for the inclusion of many-body effects. The remainder of the manuscript is structured as follows. First, we review the real-time approach \cite{Attaccalite2013} in detail (Section \ref{sc:methods}). In Section \ref{sc:our_work}, we apply Floquet theory to the problem at hand and present a Floquet formulation of the electron-field coupling operator derived from the Berry-phase polarisation. In this section, we also give details on the computational implementation of our scheme. We tested our method with several materials and benchmarked it against the real-time approach in Section \ref{sc:results}. Further discussion on the performance of our scheme and its limitations follows, before reaching the conclusions in Section \ref{sc:conclusions}.

\section{Theoretical background}\label{sc:methods}
We consider the Hamiltonian of a crystalline solid coupled to a time-dependent electric field,
\begin{equation}\label{eq:Hinitial}
    \hat {\cal H} = \hat H^{0}+ \hat H^{\cal E},
\end{equation}
where $\hat H^0$ is the zero-field unperturbed Hamiltonian while $\hat H^{\cal E}$ represents the perturbation. We denote the Bloch eigenstates of the cell-periodic unperturbed Hamiltonian, $e^{-i\kpt \cdot\mathbf{r}} \hat H^0 e^{i\kpt \cdot\mathbf{r}}$, as $\psi_{\kpt n}(\mathbf{r}) = e^{i\kpt \cdot\mathbf{r}}\; \mu_{\kpt n}(\mathbf{r})$. In what follows, the periodic part of these functions will be referred to as the zero-field time-zero states, $\tiKS$, and will be used as a starting point for time integration or as a basis.

\subsection{Real-time approach}\label{sc:methods-RT}
The real-time approach to nonlinear optics, as referred to within this manuscript, was set out by Attaccalite and Gr\"uning \cite{Attaccalite2013}, and follows the scheme presented by Souza \textit{et al.} for the Berry-phase polarisation \cite{Souza2004}. The central objects in this formalism are the time-dependent Bloch states, $\tdKS$, which represent the periodic part of the states, $\psi_{\kpt n}(\mathbf{r},t) = e^{i\kpt \cdot\mathbf{r}}\;v_{\kpt n}(\mathbf{r},t)$. The latter are obtained upon time-evolution (with an electric field) of the Bloch eigenstates of $\hat H^0_\kpt$. An EOM is then formulated for these time-dependent states as,
\begin{equation}\label{eq:EOM}
   \left(\hat H^\text{eff}_\kpt -i\partial_t\right)\tdKS = 0,
\end{equation}
with the effective Hamiltonian,
\begin{equation}\label{eq:Heff}
    \hat H^\text{eff} = \hat H^{0}+ \hat W({\cal E}).
\end{equation}
The unperturbed Hamiltonian in Eq. \ref{eq:Heff}, $\hat H^{0}$, is a single-particle operator that varies according to the level of theory considered \cite{Gruning2014,PhysRevB.94.035149}. The perturbation, $\hat W(\cal E)$, represents the coupling with the external field, ${\cal E}$, and is defined as, 
\begin{equation}\label{eq:Wdef}
    \hat W({\cal E}) = \hat w({\cal E})+ \hat{ w}^\dagger({\cal E}).
\end{equation}
In Eq. \ref{eq:Wdef}, $\hat w({\cal E})$ is the electron-field coupling operator in its Berry-phase formulation as outlined in Refs. \cite{Attaccalite2013} and \cite{Souza2004},
\begin{equation}\label{w_lowercase}
    \hat w(\mathcal{E}) = i\frac{e}{4\pi} \sum_{i=1}^3 N_i^{\parallel}\left({\cal E}\cdot {\bf a}_i\right) \sum_\sigma \sigma \hat P_{\kpt\kplus},
\end{equation}
with $\sigma = \pm 1$ and $\kpt_i^\sigma = \kpt + \sigma \Delta \kpt_i$, i.e., the \textit{next} $\kpt$-point in the grid along the $i$ Cartesian direction (the definition of \textit{next} depends on the sign of $\sigma$). The projector operator in Eq. \ref{w_lowercase} has the form,
\begin{equation}\label{eq:projector}
 \hat P_{\kpt\kplus} = \sum_{m=1}^M |\tilde v_{\kplus m} \rangle\langle v_{\kpt m}|,
\end{equation}
where $m$ runs over the occupied bands, $M$. The state $|\tilde v_{\kpt_i^\sigma m }\rangle$ is the so-called dual of the state $|v_{\kpt m} \rangle$, namely, 
\begin{equation} \label{eq:dual}
  |\tilde v_{\kplus n} \rangle = \sum_{m=1}^M \; [S_{\kpt\kplus}^{-1}]_{m,n} \;|v_{\kplus m} \rangle,
\end{equation}
with the overlap matrix elements
\begin{equation}\label{eq:Overlaps}
    [S_{ \kpt \kplus}]_{n,m} = \langle v_{\kpt n}  | v_{\kplus m}  \rangle.
\end{equation}
The real-time approach \cite{Attaccalite2013} consists on integrating the EOMs given by Eq. \ref{eq:EOM} starting from the corresponding zero-field time-zero states, $\tiKS$. This allows us to obtain the time-dependent states $\tdKS$ at every time step $t_i$, with which we can update the overlaps $[S_{ \kpt \kplus}]_{n,m}$ (Eq. \ref{eq:Overlaps}). Ultimately, we can use these overlaps to calculate the polarisation in its Berry-phase formulation,
\begin{equation}\label{Polarization_t}
    \mathbf{P}_\alpha = -\frac{e \; f}{2\pi v}\frac{\mathbf{a}_\alpha}{N_{\kpt_{\alpha}^{\perp}}} \sum_{\kpt_{\alpha}^{\perp}} \; \text{Im}\left[ \;\text{ln}\left( \prod_{i=1}^{N_{\kpt_{\alpha}}-1} \text{det} \; (\mathbf{S}_{ \kpt \kplus}) \right)\right],
\end{equation}
with the electron charge $e$, occupation factor $f$, unit cell volume $v$. Eq. \ref{Polarization_t} provides the dynamical polarisation in the direction $\alpha$ of the lattice vector $\mathbf{a}_{\alpha}$. The corresponding reciprocal lattice vector $\mathbf{b}_{\alpha}$ is used to determine the number of $\kpt$-points in a \textit{string} along its direction, $N_{\kpt_{\alpha}}$, as well as the number of $\kpt$-points in a plane perpendicular to $\mathbf{b}_{\alpha}$, namely $N_{\kpt_{\alpha}^{\perp}}$. Within the regime where the dynamical polarisation is time-periodic, it can be formulated as a Fourier series, 
\begin{equation}\label{Fourier_expansion_of_P}
\text{P}(t) = \sum_{n}{\text{p}^{(n)} e^{i n \omega_0 t}},
\end{equation}
where scalar magnitudes are used for simplicity. In addition, one can consider its expansion in orders of the electric field ${\cal E}$ \cite{Boyd2008},
\begin{equation}\label{eq:expansion_in_E}
    \text{P}(t) = \chi^{(1)} \mathcal{E}(t) + \chi^{(2)} \mathcal{E}^2(t) + \chi^{(3)} \mathcal{E}^3(t) + \mathcal{O}(\mathcal{E}^4(t)),
\end{equation}
where the tensor nature of the susceptibilities $\chi^{(n)}$ is omitted for brevity. Comparing Eqs. \ref{Fourier_expansion_of_P} and \ref{eq:expansion_in_E} finally allows us to extract susceptibilities to any order. The relation between the Fourier coefficients, $\text{p}^{(n)}$, and the desired susceptibilities will depend on the order, $n$, and the shape of the electric field, which would typically be a sine function ($\frac{e^{i\omega_0t} - e^{-i\omega_0t}}{2i}$).

As mentioned in the introduction, this framework has two main advantages. First, as a non-perturbative scheme, it allows for the simultaneous determination of susceptibilities to different orders in the electric field. This is also facilitated by having a Berry-phase derived electron-field coupling operator that remains valid to every order in the electric field. Second, the inclusion of many-body effects is as simple as adding terms to the effective Hamiltonian used in the EOM, Eq. \ref{eq:EOM} \cite{Gruning2014}. 

Despite its many virtues, the real-time approach often presents challenges regarding its elevated computational cost. This is its biggest disadvantage and originates from its time-propagation nature. Arguably, there is a particular case in which much of this cost is avoidable, i.e., computing nonlinear optical susceptibilities. In these calculations, the system is driven by a periodic perturbation and the response is sampled at a handful of times within one period, i.e., only one period worth of dynamical polarisation data is needed to extract susceptibilities. However, a considerably longer time is required to \textit{dephase} the response before sampling it, in order to filter out all the eigenfrequencies that are excited when the electric field is first introduced. This amounts to a total simulated time that greatly exceeds the time window actually used to probe the response. This long simulated time combined with the expensive numerical integration of the EOMs (often with short time steps) render this kind of calculations particularly costly. It would then be desirable to devise a strategy where the dephasing is not needed, numerical time-evolution is avoided and/or the problem becomes time-independent altogether. We will see in Section \ref{sc:our_work} how Floquet theory offers a framework in which all of the above are possible.

\section{Nonlinear optics \textit{via} Floquet theory and Berry-phase polarisation}\label{sc:our_work}
In this section, we apply Floquet theory to the EOM (Eq. \ref{eq:EOM}) of the real-time approach \cite{Attaccalite2013}. We start by defining the so-called Floquet-Kohn-Sham (FKS) basis in Section \ref{sc:FKSbasis}. This allows us to turn Eq. \ref{eq:EOM} into a self-consistent time-independent eigenproblem (Section \ref{sc:QEproblem}). In Section \ref{sc:W_in_FKS}, we derive an expression for the electron-field coupling operator in FKS basis. We use its Berry-phase formulation \cite{Attaccalite2013,Souza2004}, which makes our approach valid for extended systems and distinguishes it from previous Floquet works \cite{Telnov1997}. Finally, we describe the computational implementation of our method in Section \ref{sc:comp-imp}.

Within this manuscript, we will choose the IPA to be defined at the density functional theory (DFT) level plus a static quasi-particle correction. With this assumption, the effective Hamiltonian in Eq. \ref{eq:Heff} takes the form, 
\begin{equation}\label{eq:Heff-IPA}
    \hat H^\text{IPA} = \underbrace{\hat H^{KS} +\hat \Delta_\text{QP}}_{\hat H^{\text{IPA},0}} \;+ \;\hat W({\cal E}) ,
\end{equation}
where $\hat H^{KS}$ represents the Kohn-Sham (KS) Hamiltonian and $\hat H^{\text{IPA},0}$ is time-independent. While the quasi-particle correction, $\Delta_\text{QP}$, could be determined by a $GW$ calculation, we only considered a rigid shift to the band structure. Since the unperturbed Hamiltonian, $\hat H^{0}$, is formulated at the DFT level, we can also refer to the zero-field states, $\tiKS$, simply as KS states.

\subsection{Time independent Floquet-Kohn-Sham basis}\label{sc:FKSbasis}
Let us assume that the effective Hamiltonian in Eq. \ref{eq:EOM} is time-periodic with a period, $T=\frac{2\pi}{\omega_0}$, given by the frequency of the perturbing electric field, $\omega_0$. Invoking Floquet's theorem, one can assert  this EOM will admit solutions in the form of the so-called Floquet basis functions, i.e., $ e^{-i\xi_\alpha t} \; \phi_\alpha(t)$, where $\xi_\alpha$ is the so-called Floquet quasi-energy and the time-dependent Floquet states, $\phi_\alpha(t)$, retain the periodicity of the Hamiltonian, $\phi_\alpha(t)=\phi_\alpha(t+T)$. The general solution to Eq. \ref{eq:EOM} would then be a linear combination of said Floquet functions,
\begin{equation}
    \label{td-FloquetBasis}
    \tdKS = \sum_\alpha{c_{\kpt n}^\alpha \; e^{-i\xi_\alpha t} \; |\phi_\alpha\rangle}.
\end{equation}
Making use of the adiabatic approximation for weak fields \cite{Kapoor2013}, we can assume each time-zero KS state will evolve adiabatically into a single time-periodic Floquet state and retain only one term in the summation in Eq. \ref{td-FloquetBasis}, i.e.,
\begin{equation}
    \label{adiabatic_approx}
    \tdKS \approx e^{-i\xi_\alpha t} \; |\phi_\alpha\rangle = e^{-i\xi_{\kpt n} t} \; \tdFL,
\end{equation}
where we replaced the label $\alpha$ with the index of the state at $\kpt$-point $\kpt$ and band $n$. 
Projecting over the zero-field KS states, $\tiKS$, we get, 
\begin{equation}
     \label{td-coefficients}
     \tdKS  = e^{-i\xi_{\kpt n} t} \; \sum_i^{+\infty}{|\mu_{\mathbf{k}i}\rangle\langle\mu_{\mathbf{k}i}} \tdFL = e^{-i\xi_{\kpt n} t} \; \sum_i^{+\infty}{d_{\kpt ni}(t) \; |\mu_{\mathbf{k}i}\rangle},
\end{equation}
where the index $i$ runs over both occupied and empty bands. As the coefficients $d_{\kpt ni}(t) \equiv\langle\mu_{\mathbf{k}i} \tdFL$ retain the time periodicity of the Floquet states, $d_{\kpt ni}(t)=d_{\kpt ni}(t+T)$, they can be expanded in a Fourier series, 
\begin{equation}
    \label{time-to-eta}
    d_{\kpt ni}(t) = \sum_{\eta=-\infty}^{+\infty}{ e^{-i\eta\omega_0t} \; \dFKS},
\end{equation}
where $\eta$ will be referred to as the Floquet mode.
Finally, with Eqs. \ref{td-coefficients} and \ref{time-to-eta}, we arrive at the representation of the time-dependent Bloch states we will use in this work,
\begin{equation}
    \label{ti-coefficients}
    \tdKS = e^{-i\xi_{\kpt n} t} \sum_{\eta=-\infty}^{+\infty}{ e^{-i\eta\omega_0t} \; \sum_i^{+\infty}{\dFKS \; |\mu_{\mathbf{k}i}\rangle}},
\end{equation}
where the coefficients $\dFKS$ depend on the band index $i$ and the Floquet mode $\eta$.
The states given by $|\kpt n i;\eta\rangle \equiv e^{-i\xi_{\kpt n} t} \; e^{-i\eta\omega_0t} \; |\mu_{\mathbf{k}i}\rangle $ form what we will refer to as FKS space. This extended Hilbert space includes $\mathcal{L}_2[0,T]$ plus the space spanned by KS eigenvectors, i.e., $\mathcal{H}$. The inner product in $\mathcal{L}_2[0,T]\otimes\mathcal{H}$ is defined as $\langle\hspace{-1.37pt}\langle\cdot|\cdot\rangle\hspace{-1.37pt}\rangle\equiv\int_0^T{dt\; \langle\cdot|\cdot\rangle}$, with $\langle\cdot|\cdot\rangle$ the usual inner product in $\mathcal{H}$. The dimension of FKS space must be truncated to $N_b \times (2\eta_{\text{max}}+1)$ for any practical calculation. $N_b$ is the number of bands used for the expansion in Eq. \ref{td-coefficients} while $\eta_{\text{max}}$ is the maximum Floquet mode used in the expansion in Eq. \ref{time-to-eta}. We note that, within this manuscript, we define $\eta_{\text{max}}$ in relation to the FKS states, i.e., $\eta_{\text{max}}$ implies the definition,
\begin{equation}
\label{eq:eta_max_condition}
%\dFKS = 0 \;\; \forall \; \eta \; \text{such that} \; |\eta| > \eta_{\text{max}}.    
\dFKS \equiv 0 \;\; \text{if} \; |\eta| > \eta_{\text{max}}.
\end{equation}

\subsection{Quasi-energy eigenproblem}\label{sc:QEproblem}
Choosing the Hamiltonian in Eq. \ref{eq:EOM} at the IPA level (Eq. \ref{eq:Heff-IPA}) and expanding the time-dependent Bloch states in FKS basis (Eq. \ref{ti-coefficients}), we arrive at the EOM,
\begin{equation}\label{eq:EOM-IPA2}
   \sum_j^{+\infty}\sum_{\gamma=-\infty}^{+\infty}\left(\hat H^\text{IPA}_\kpt -i\partial_t\right)e^{-i\xi_{\kpt n} t} { e^{-i\gamma\omega_0t} \; {\widetilde d_{\kpt nj} (\gamma) |\mu_{\mathbf{k}j}\rangle}} = 0.
\end{equation}
Acting the operators $\hat H^0$ and $-i\partial_t$ we obtain,
\begin{equation}\label{eq:EOM-IPA3}
   \sum_j^{+\infty}\sum_{\gamma=-\infty}^{+\infty}\left( E^\text{IPA}_{\kpt j} + \hat W_{\kpt}(t) - \xi_{\kpt n} -\gamma\omega_0 \right)e^{-i\xi_{\kpt n} t} { e^{-i\gamma\omega_0t} \; {\widetilde d_{\kpt nj} (\gamma) |\mu_{\mathbf{k}j}\rangle}} = 0,
\end{equation}
where $E^\text{IPA}$ are the KS energies shifted by the quasi-particle corrections and the time-dependence of $\hat W_{\kpt}(\cal E)$ is shown explicitly. We multiply to the left by $\int dt \; e^{+i\xi_{\kpt n} t}\; e^{+i\eta\omega_0 t}\; \langle \mu_{\kpt i}|$ and arrive at
\begin{equation}\label{eq:EOM-IPA4}
   \sum_j^{+\infty}\sum_{\gamma=-\infty}^{+\infty} \Bigl[ \left( E^\text{IPA}_{\kpt j}  - \xi_{\kpt n} -\gamma\omega_0 \right) \delta_{i,j} \delta_{\eta,\gamma} \; + W_{\kpt ij}(\eta,\gamma) \Bigr]\; \widetilde d_{\kpt nj} (\gamma) = 0,
\end{equation}
where $W_{\kpt ij}(\eta,\gamma)$ are the matrix elements of $\hat W_{\kpt}(t)$ in FKS space. It is also worth noting that the operator $\left(\hat H_{\kpt}^0 -i\partial_t\right)$ is diagonal in FKS space, with matrix elements given by $\left(E^{\text{IPA}}_{\kpt j}-\xi_{\kpt n}-\gamma\omega_0\right)$. Now, Eq. \ref{eq:EOM-IPA4} can be rearranged as an eigenvalue problem for the Floquet quasi-energies,  
\begin{equation}\label{eq:Eigenproblem}
   \sum_j^{+\infty}\sum_{\gamma=-\infty}^{+\infty} \Bigl[ \left( E^\text{IPA}_{\kpt j}  -\gamma\omega_0 \right) \delta_{i,j} \delta_{\eta,\gamma} \; + W_{\kpt ij}(\eta,\gamma) \Bigr]\; \widetilde d_{\kpt nj} (\gamma) = \xi_{\kpt n} \; \dFKS,
\end{equation}
We define the operator on the LHS of Eq. \ref{eq:Eigenproblem} as the quasi-energy operator $\hat {\cal K}_{\kpt}$. Its matrix elements in FKS space are,
\begin{equation}\label{eq:QEoperator}
     {\cal K}_{\kpt ij}\left(\eta,\gamma\right) = \left( E^\text{IPA}_{\kpt j}  -\gamma\omega_0 \right) \delta_{i,j} \delta_{\eta,\gamma} \; + W_{\kpt ij}(\eta,\gamma),
\end{equation}
and the eigenvalue problem in Eq. \ref{eq:Eigenproblem} reduces to the shorthand notation,
\begin{equation}\label{eq:Eigenproblem_2}
   \sum_j^{+\infty}\sum_{\gamma=-\infty}^{+\infty} {\cal K}_{\kpt ij}\left(\eta,\gamma\right)\; \widetilde d_{\kpt nj} (\gamma) = \xi_{\kpt n} \; \dFKS,
\end{equation}

Formally, the matrix elements $W_{\kpt ij}(\eta,\gamma)$ can be obtained by expressing the time-periodic electron-field coupling operator in Floquet space, i.e., $\hat W_{\kpt}(t) = \sum_{\zeta=-\infty}^{+\infty} e^{-i\omega_0\zeta t}\; \tilde W_{\kpt}(\zeta)$, and taking the inner product,
\begin{equation}\label{eq:Wmatrixelement_1}
   W_{\kpt ij}(\eta,\gamma) = \int dt\; e^{+i\omega_0\eta t}\; \sum_{\zeta=-\infty}^{+\infty} e^{-i\omega_0\zeta t}\;e^{-i\omega_0\gamma t}\;\langle\mu_{\kpt i}| \tilde W_{\kpt}(\zeta)|\mu_{\kpt j}\rangle.
\end{equation}
Replacing the time integral with a delta function, we arrive at,
\begin{eqnarray}\label{eq:Wmatrixelement_2}
   W_{\kpt ij}(\eta,\gamma) &= \sum_{\zeta=-\infty}^{+\infty} \delta_{\zeta,\eta-\gamma}\;\langle\mu_{\kpt i}| \tilde W_{\kpt}(\zeta)|\mu_{\kpt j}\rangle \nonumber\\
   &= \langle\mu_{\kpt i}| \tilde W_{\kpt}(\eta-\gamma)|\mu_{\kpt j}\rangle.
\end{eqnarray}
An expression for $W_{\kpt ij}(\eta,\gamma)$ will be obtained in Section \ref{sc:W_in_FKS}. We can anticipate from Eq. \ref{eq:Wmatrixelement_2} that $W_{\kpt ij}(\eta,\gamma)$ will couple different Floquet modes in the eigenproblem of Eq. \ref{eq:Eigenproblem}.

\subsection{Electron-field coupling operator $\hat W(\cal E)$}\label{sc:W_in_FKS}
In order to obtain the matrix elements $W_{\kpt ij}(\eta,\gamma)$ (Eq. \ref{eq:Wmatrixelement_2}), we consider a sinusoidal electric field of amplitude ${\cal E}_0$ and frequency $\omega_0$, and re-write Eq. \ref{w_lowercase} as, 
\begin{eqnarray}\label{w_k}
    \hat w_\kpt &= i\frac{e}{4\pi} \sum_{i=1}^3 N_i^{\parallel}\left({\cal E}_0\cdot {\bf a}_i\right) \sum_\sigma \sigma \left( \frac{e^{i\omega_0t} - e^{-i\omega_0t}}{2i} \right) \hat P_{\kpt\kplus} \nonumber\\
    &= \frac{e}{8\pi} \sum_{i=1}^3 N_i^{\parallel}\left({\cal E}_0\cdot {\bf a}_i\right) \sum_\sigma \sigma \sum_{\sigma_2} \sigma_2\; e^{\sigma_2 \;i\omega_0t} \hat P_{\kpt\kplus},
\end{eqnarray}
where $\sigma_2=\pm1$. The summations in Eq. \ref{w_k} add up to twelve equivalent terms. In what follows, we will work with just one of them for simplicity. Choosing the positive exponential ($\sigma_2=+1$) and replacing $\kplus$ by $\kpt^+$, we define,  
\begin{equation}\label{eq:projector_k}
    \hat {\cal P}^+_{\kpt\kpt^+} = e^{+i\omega_0t} \hat P_{\kpt\kpt^+}.
\end{equation}
We start by acting $\hat{\cal P}^+_{\kpt\kpt^+}$ on a time-dependent Bloch state,
\begin{equation}\label{eq:Wderivation_1}
    \hat {\cal P}^+_{\kpt\kpt^+} \; \tdKS = \hat {\cal P}^+_{\kpt\kpt^+} \; e^{-i\xi_{\kpt n} t} \sum_{\gamma=-\infty}^{+\infty}{ e^{-i\gamma\omega_0t} \; \sum_j^{+\infty}{\dFKSj |\mu_{\mathbf{k}j}\rangle}},
\end{equation}
where we have expanded the state $\tdKS$ in FKS basis as shown in Eq. \ref{ti-coefficients}.
Multiplying the RHS of Eq. \ref{eq:Wderivation_1} to the left by $\int dt \; e^{+i\xi_{\kpt n} t}\; e^{+i\eta\omega_0 t}\; \langle \mu_{\kpt i}|$, we arrive at
\begin{eqnarray}\label{Wderivation_2}
   \sum_{\gamma=-\infty}^{+\infty} \sum_j^{+\infty} \int dt\;&e^{+i\eta\omega_0t} \;\langle \mu_{\kpt i}|\; \hat {\cal P}^+_{\kpt\kpt^+} \; { |\mu_{\mathbf{k}j}\rangle} \; e^{-i\gamma\omega_0t} \; \dFKSj \nonumber
   \\
   &=\sum_{\gamma=-\infty}^{+\infty}\sum_j^{+\infty}{\cal P}^+_{\kpt\kpt^+ij}(\eta,\gamma)\; \dFKSj.
\end{eqnarray}
In Eq. \ref{Wderivation_2}, we have extracted an expression for the matrix elements of the projector,
\begin{equation}\label{Wderivation_3}
   {\cal P}^+_{\kpt\kpt^+ij}(\eta,\gamma) = \int dt\; e^{+i\eta\omega_0t} \;\langle \mu_{\kpt i}|\; \hat {\cal P}_{\kpt\kpt^+} \; { |\mu_{\mathbf{k}j}\rangle} \; e^{-i\gamma\omega_0t}.
\end{equation}
Using Eqs. \ref{eq:projector} and \ref{eq:projector_k}, we find
\begin{equation}\label{Wderivation_4}
   {\cal P}^+_{\kpt\kpt^+ij}(\eta,\gamma) = \int dt\; e^{+i(\eta+1)\omega_0t} \;\langle \mu_{\kpt i}| \sum_{m}^M |\tilde v_{\kpt^+ m} \rangle\langle v_{\kpt m} 
    |\mu_{\mathbf{k}j}\rangle \; e^{-i\gamma\omega_0t}.
\end{equation}
The dual $|\tilde v_{\kpt^+ m} \rangle$ can be expressed via the overlaps matrix as in Eq. \ref{eq:dual},
\begin{equation}\label{Wderivation_5}
\fl   {\cal P}^+_{\kpt\kpt^+ij}(\eta,\gamma) = \int dt\; e^{+i(\eta+1)\omega_0t} \;\langle \mu_{\kpt i}|\sum_{m}^M \sum_{m_1}^M [S_{\kpt\kpt^+}^{-1}]_{m_1,m}\;|v_{\kpt^+ m_1} \rangle \langle v_{\kpt m}|\mu_{\mathbf{k}j}\rangle \; e^{-i\gamma\omega_0t}.
\end{equation}
Eq. \ref{Wderivation_5} contains the inverse of the matrix $\mathbf{S}_{\kpt\kpt^+}$, i.e., the time-dependent overlap already defined in Eq. \ref{eq:Overlaps}. Transforming Eq. \ref{eq:Overlaps} to FKS basis (Eq. \ref{ti-coefficients}), we arrive at,
\begin{eqnarray}\label{Sderivation_1}
    [S_{\kpt\kpt^+}]_{m,m_1} & = \;
    e^{+i\xi_{\kpt m} t}\;  
    e^{-i\xi_{\kpt^+ m_1} t} 
    \sum_{\zeta_2=-\infty}^{+\infty}
    \sum_{\eta_2=-\infty}^{+\infty}
    e^{+i\zeta_2\omega_0t}
    e^{-i\eta_2\omega_0t}\; \times \nonumber \\
    &\sum_{j_2}^{+\infty} 
    \sum_{i_2}^{+\infty}
    \widetilde d^*_{\kpt mj_2}(\zeta_2) \langle\mu_{\kpt j_2}|\mu_{\kpt^+ i_2}\rangle d_{\kpt^+ m_1i_2}(\eta_2).
\end{eqnarray}
Defining the time-zero zero-field overlaps as $[S_{\kpt\kpt^+}^0]_{j_2,i_2} = \langle\mu_{\kpt j_2}|\mu_{\kpt^+ i_2}\rangle$ and making the replacement $\eta_2' = \eta_2-\zeta_2$, we get to,
\begin{eqnarray}\label{Sderivation_2}
    [S_{\kpt\kpt^+}]_{m,m_1} & =
    e^{+i\xi_{\kpt m} t}\;  
    e^{-i\xi_{\kpt^+ m_1} t} 
    \sum_{\eta_2'=-\infty}^{+\infty}
    e^{-i\eta_2'\omega_0t}\;\times\nonumber\\
    &\left(\sum_{\zeta_2=-\infty}^{+\infty}
    \sum_{j_2}^{+\infty} 
    \sum_{i_2}^{+\infty}
    \widetilde d^*_{\kpt mj_2}(\zeta_2) [S_{\kpt\kpt^+}^0]_{j_2,i_2} d_{\kpt^+ m_1i_2}(\eta_2'+\zeta_2)\right).
\end{eqnarray}
We now define the terms in the parentheses as $[\widetilde S_{\kpt\kpt^+}]_{m,m_1}(\eta_2')$ and obtain,
\begin{equation}\label{Sderivation_3}
    [S_{\kpt\kpt^+}]_{m,m_1} =
    e^{+i\xi_{\kpt m} t}\;  
    e^{-i\xi_{\kpt^+ m_1} t} 
    \sum_{\eta_2'=-\infty}^{+\infty}
    e^{-i\eta_2'\omega_0t} [\widetilde S_{\kpt\kpt^+}]_{m,m_1}(\eta_2'),
\end{equation}
which implies the need for a self-consistent solution to the eigenproblem in Eq. \ref{eq:Eigenproblem}, given that the matrix elements $W_{\kpt ij}(\eta,\gamma)$ will depend on the solutions $\{\dFKS\}_{\kpt,n,i,\eta}$ through the overlap matrices $[S_{\kpt\kpt^+}]_{m,m_1}$.

According to Eq. \ref{Wderivation_5}, we need the inverse of the matrix $\mathbf{S}_{\kpt\kpt^+}$. How this inversion is performed will be discussed in Section \ref{sc:comp-imp}. For now, we will assume we can find an expression for $[S_{\kpt\kpt^+}^{-1}]_{m_1,m}$ expanded in Floquet modes as,
\begin{equation}\label{Sderivation_4}
    [S_{\kpt\kpt^+}^{-1}]_{m_1,m} = 
    e^{-i\xi_{\kpt m} t}\;  
    e^{+i\xi_{\kpt^+ m_1} t} 
    \sum_{\eta_2'=-\infty}^{+\infty}
    e^{-i\eta_2'\omega_0t} [\widetilde D_{\kpt\kpt^+}]_{m_1,m}(\eta_2'),
\end{equation}
where the signs of the quasi-energy exponentials have been inverted with respect to Eq. \ref{Sderivation_3}.
Inserting Eq. \ref{Sderivation_4} into Eq. \ref{Wderivation_5}, we arrive at,
\begin{eqnarray}\label{Wderivation_6}
   &{\cal P}^+_{\kpt\kpt^+ij}(\eta,\gamma) = \int dt\; e^{-i(\gamma-\eta-1)\omega_0t} \sum_{m}^M \sum_{m_1}^M e^{-i\xi_{\kpt m} t}\;  
    e^{+i\xi_{\kpt^+ m_1} t} \;\times\nonumber \\
   &\langle \mu_{\kpt i}|\left(\; \sum_{\eta_2'=-\infty}^{+\infty}
    e^{-i\eta_2'\omega_0t} [\widetilde D_{\kpt\kpt^+}]_{m_1,m}(\eta_2') |v_{\kpt^+ m_1} \rangle \langle v_{\kpt m}|
   \; \right)|\mu_{\mathbf{k}j}\rangle.
\end{eqnarray}
Furthermore, we now use Eq. \ref{ti-coefficients} to replace the time-dependent Bloch states in Eq. \ref{Wderivation_6} as,
\begin{equation}\label{Wderivation_7}
    \langle v_{\kpt m}| = e^{+i\xi_{\kpt m} t} \sum_{\zeta_1=-\infty}^{+\infty} e^{+i\zeta_1\omega_0t} \; \sum_{i_3}^{+\infty}\widetilde d^*_{\kpt mi_3} (\zeta_1) \langle\mu_{\kpt i_3}|
\end{equation}
and 
\begin{equation}\label{Wderivation_8}
    |v_{\kpt^+ m_1}\rangle = e^{-i\xi_{\kpt^+ m_1} t} \sum_{\eta_1=-\infty}^{+\infty} e^{-i\eta_1\omega_0t} \; \sum_{i_1}^{+\infty}\widetilde d_{\kpt^+ m_1i_1} (\eta_1) |\mu_{\kpt^+ i_1}\rangle.
\end{equation}
Inserting Eqs. \ref{Wderivation_7} and \ref{Wderivation_8} into Eq. \ref{Wderivation_6}, the quasi-energy exponentials cancel each other out and we arrive at
\begin{eqnarray}\label{Wderivation_9}
\fl   {\cal P}^+_{\kpt\kpt^+ij}(\eta,\gamma) = \int dt\; e^{-i(\gamma-\eta-1)\omega_0t} \sum_{m}^M \sum_{m_1}^M  \sum_{\eta_2'=-\infty}^{+\infty}
    e^{-i\eta_2'\omega_0t} [\widetilde D_{\kpt\kpt^+}]_{m_1,m}(\eta_2')\; \times\nonumber\\
\fl   \sum_{\eta_1=-\infty}^{+\infty} e^{-i\eta_1\omega_0t} \; \sum_{i_1}^{+\infty}\widetilde d_{\kpt^+ m_1i_1} (\eta_1) \langle \mu_{\kpt i}|\mu_{\kpt^+ i_1}\rangle 
   \sum_{\zeta_1=-\infty}^{+\infty} e^{+i\zeta_1\omega_0t} \; \sum_{i_3}^{+\infty}\widetilde d^*_{\kpt mi_3} (\zeta_1) \langle\mu_{\kpt i_3}|\mu_{\mathbf{k}j}\rangle,
\end{eqnarray}
where $\langle\mu_{\kpt i_3}|\mu_{\mathbf{k}j}\rangle = \delta_{i_3,j}$ eliminates the summation over $i_3$ and a new zero-field overlap is formed, namely $[S_{\kpt\kpt^+}^0]_{i,i_1} =\langle \mu_{\kpt i}|\mu_{\kpt^+ i_1}\rangle$. Grouping all the Floquet mode summations and exponentials together, we obtain
\begin{eqnarray}\label{Wderivation_10}
   &{\cal P}^+_{\kpt\kpt^+ij}(\eta,\gamma) = \sum_{\eta_2'=-\infty}^{+\infty}
   \sum_{\eta_1=-\infty}^{+\infty}
   \sum_{\zeta_1=-\infty}^{+\infty}
   \left(\int dt\; e^{-i(\gamma-\eta-1+\eta_2'+\eta_1-\zeta_1)\omega_0t}\right)\; \times\nonumber\\
   &\sum_{m}^M \sum_{m_1}^M \; 
   [\widetilde D_{\kpt\kpt^+}]_{m_1,m}(\eta_2')
   \; \sum_{i_1}^{+\infty}\;\widetilde d_{\kpt^+ m_1i_1} (\eta_1) [S_{\kpt\kpt^+}^0]_{i,i_1} 
   \; \widetilde d^*_{\kpt mj} (\zeta_1) ,
\end{eqnarray}
where the first parenthesis encapsulates the time dependence. This term results in the condition $(\gamma-\eta-1+\eta_2'+\eta_1-\zeta_1) = 0$. Choosing to replace $\zeta_1$ and thus eliminate this summation via $\delta_{\zeta_1,\gamma-\eta-1+\eta_2'+\eta_1}$, we finally arrive at
\begin{eqnarray}\label{Wderivation_11}
   {\cal P}^+_{\kpt\kpt^+ij}(\eta,\gamma) =  \sum_{m}^M &\sum_{\eta_2'=-\infty}^{+\infty}
   \sum_{\eta_1=-\infty}^{+\infty} \widetilde d^*_{\kpt mj} (\gamma-\eta-1+\eta_2'+\eta_1)\; \times\nonumber\\
   &\sum_{m_1}^M \;
   [\widetilde D_{\kpt\kpt^+}]_{m_1,m}(\eta_2')
   \; \sum_{i_1}^{+\infty}  [S_{\kpt\kpt^+}^0]_{i,i_1} \widetilde d_{\kpt^+ m_1i_1}(\eta_1).
\end{eqnarray}
The expressions for the eleven remaining instances of $\hat{\cal P}^{\sigma_2}_{\kpt\kpt_i^\sigma}$ can be derived by analogy to Eq. \ref{Wderivation_11}. Once all these projectors are computed, we can go back to Eqs. \ref{w_k}, \ref{eq:projector_k} and \ref{eq:Wdef} to finally obtain the matrix elements of the electron-field coupling operator $W_{\kpt ij}(\eta,\gamma)$ in FKS space. 

To summarise, we have laid out all the steps needed to reformulate the time-dependent real-time EOM (Eq. \ref{eq:EOM}) into a time-independent eigenproblem (Eq. \ref{eq:Eigenproblem}) in FKS basis (Eq. \ref{ti-coefficients}). Crucially, our scheme is valid for extended systems since we use the Berry-phase derived electron-field coupling operator. The latter depends on the solutions $\{\dFKS\}_{\kpt,n,i,\eta}$ and thus the eigenproblem in Eq. \ref{eq:Eigenproblem} must be solved self-consistently. As our Floquet reformulation is time-independent, it does not require dephasing or expensive numerical time-integrations, which will alleviate the computational burden. Nonetheless, we retain the main advantage of the real-time approach, i.e., the scheme remains non-perturbative in the electric field, allowing for the simultaneous calculation of susceptibilities to different orders in the electric field.

\subsection{Computational implementation}\label{sc:comp-imp}
The Floquet scheme presented here involves, for each frequency $\omega_0$, a self-consistency cycle where the eigenvectors calculated in one iteration (solving Eq. \ref{eq:Eigenproblem}) are fed to the next one (Eqs. \ref{Sderivation_2} and \ref{Wderivation_11}) until convergence is reached. The condition for convergence is based on the absolute error in the real and imaginary parts of the susceptibility to every order requested by the user. As the real-time approach \cite{Attaccalite2013}, this Floquet implementation offers parallelisation in frequencies and $\kpt$-points. Moreover, it works both in the non-magnetic (spin unpolarised) and magnetic non-collinear (spinorial) formulations. 
One of the challenges met during this implementation concerns the inversion of the overlap matrix $\mathbf{S}_{\kpt\kpt^+}$, which is required to obtain the coefficients $[\widetilde D_{\kpt\kpt^+}]_{m_1,m}(\eta_2')$ as defined in Eq. \ref{Sderivation_4}. To this end, two strategies were implemented and compared. The first one would entail avoiding the time domain entirely and remaining in Floquet space, i.e., one could obtain the coefficients $[\widetilde D_{\kpt\kpt^+}]_{m_1,m}(\eta_2')$ (see Eq. \ref{Sderivation_4}) directly from the coefficients $[\widetilde S_{\kpt\kpt^+}]_{m_1,m}(\eta_2')$ (see Eq. \ref{Sderivation_3}). This is indeed possible for scalar functions \cite{Duffin1962} and was extended to matrices. The alternative option is to trivially go to the time domain evaluating Eq. \ref{Sderivation_3} for several sample times $t_i$, invert the matrices $[S_{\kpt\kpt^+}]_{m_1,m}(t_i)$ numerically at each $t_i$, and Fourier transform the resulting $[S_{\kpt\kpt^+}^{-1}]_{m_1,m}(t_i)$ back to Floquet space, i.e., solve Eq. \ref{Sderivation_4} for $[\widetilde D_{\kpt\kpt^+}]_{m_1,m}(\eta_2')$. This numerical inversion in time domain resulted both more robust and less time-consuming than the approach based on Duffin's theorems \cite{Duffin1962}. 

The calculation of the polarisation \textit{via} Eq. \ref{Polarization_t} proved to be another obstacle in our implementation. As before, the problem originates from the fact that what is available to us are the Floquet coefficients of each overlap matrix, $[\widetilde S_{\kpt\kpt^+}]_{m_1,m}(\eta_2')$, rather than the matrix itself, $[S_{\kpt\kpt^+}]_{m_1,m}$. Inserting Eq. \ref{Sderivation_3} into Eq. \ref{Polarization_t}, we see that we would need to calculate the logarithm of a sum, which should be linearised by expanding it into a logarithmic series if the $\eta'_2=0$ term dominates. While this does indeed lead to a manageable set of equations to solve, it would turn our scheme into a perturbative one, thus losing one of the great advantages of the real-time approach. In fact, the issue of what order in this perturbative expansion corresponds to which order of the response in the electric field does not seem to be a trivial one. As before, the alternative is to switch to the time domain, i.e., evaluate Eq. \ref{Sderivation_3} for a handful of sample times $t_i$, calculate the polarisation at each $t_i$ with Eq. \ref{Polarization_t} and proceed with the usual steps in Eq. \ref{eq:expansion_in_E} to extract the required susceptibilities.

This resembles the usual choice one has in systems with translational invariance of going back and forth from real to reciprocal space to calculate whatever operator is simpler in either basis. By analogy, we can switch to time domain to perform a series of operations and then Fourier transform back to Floquet space. It is key to understand that transforming to the time domain does not necessarily imply the long simulated times and short time steps (i.e., tens of thousands of sample times) characteristic of the real-time approach to nonlinear optics. Rather, the assumed time periodicity means that one just needs to calculate the observables across one time period only. Moreover, the number of time steps required within that period is very limited as it corresponds to the total number of Floquet modes one needs for $[\widetilde D_{\kpt\kpt^+}]_{m_1,m}(\eta_2')$, which happens to be ($2\times2\eta_{\text{max}}+1$), i.e., $\eta_{\text{max}}^{S} = 2\eta_{\text{max}}$.

Finally, we introduced dissipation effects as a phenomenological damping term in the diagonal of the quasi-energy operator,
\begin{equation}\label{eq:Damping}
   {\cal K}_{\kpt ij} \left(\eta,\gamma\right) = \left( E^\text{IPA}_{\kpt j}  -\gamma\omega_0 - \underbrace{i\nu_d\left(1-\delta_{\gamma,0}\right)}_\text{Damping}\right)\delta_{i,j} \delta_{\eta,\gamma} + W_{\kpt ij}\left(\eta,\gamma\right),
\end{equation}
where $\nu_d$ is a positive real number that provides the broadening to the spectra. This damping term deals with the rich structure of avoided crossings characteristic of Floquet quasi-energy operators, effectively removing the singularities that appear at resonant frequencies. The term $\left(1-\delta_{\gamma,0}\right)$ ensures that processes to all orders are damped to the same extent. We also introduced a small imaginary contribution to some KS eigenenergies (typically $\nu_d\times10^{-4}$) in order to avoid singularities arising from degeneracies or crossings in the KS band structure. This applies to those eigenvalues that differ in less than, e.g., $1\times10^{-7}$ Ha, and is only added at the $0^{th}$ Floquet mode. Due to the introduction of these imaginary contributions, the quasi-energy matrix is no longer Hermitian. We used a diagonalisation routine suitable for non-Hermitian matrices and verified that the imaginary part of the Floquet quasi-energies remained negligible.
 
\section{Results and Discussion}\label{sc:results}
The Floquet approach to nonlinear optics presented in this manuscript has been implemented into the Yambo code \cite{Marini2009,Sangalli2019} and tested with a number of well-known materials. The latter have been studied before from an \textit{ab-initio} perspective \cite{Luppi2010,Margulis2013,Trolle2014} and within the real-time formalism in particular \cite{Attaccalite2013,Gruning2014}, which makes them ideal for validating and benchmarking our method. To this end, a systematic comparison between the real-time \cite{Attaccalite2013} and Floquet approaches has been conducted, where the real-time calculations were also performed using the Yambo code \cite{Marini2009,Sangalli2019}. The starting KS wavefunctions and energies were computed with Quantum Espresso \cite{Giannozzi2017}.

\subsection{Second Harmonic Generation}\label{sc:SHG}
\begin{figure}
    \centering
    \includegraphics{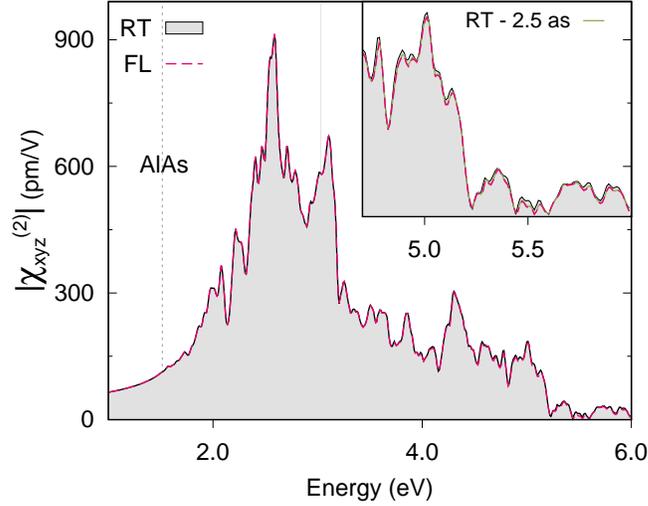}
    \caption{Bulk AlAs SHG spectra on a $30\times30\times30$ $\kpt$-grid with a broadening of \SI{.04}{\eV}, calculated by the real-time (RT) approach (black line with grey filling) and our Floquet (FL) method (pink dashed line). The intensity of the electric field is 1$\times 10^3$ \si{\kilo\watt\per\centi\meter\squared}. The real-time spectrum is calculated with a 10-\si{\atto\second} time step. The inset shows a portion of the spectrum re-calculated with a 1-\si{\atto\second} time step (green line), which shows better agreement with the Floquet spectrum.}
    \label{fig:SHG_AlAs}
\end{figure}

\begin{figure}[h!]
    \centering
    \includegraphics{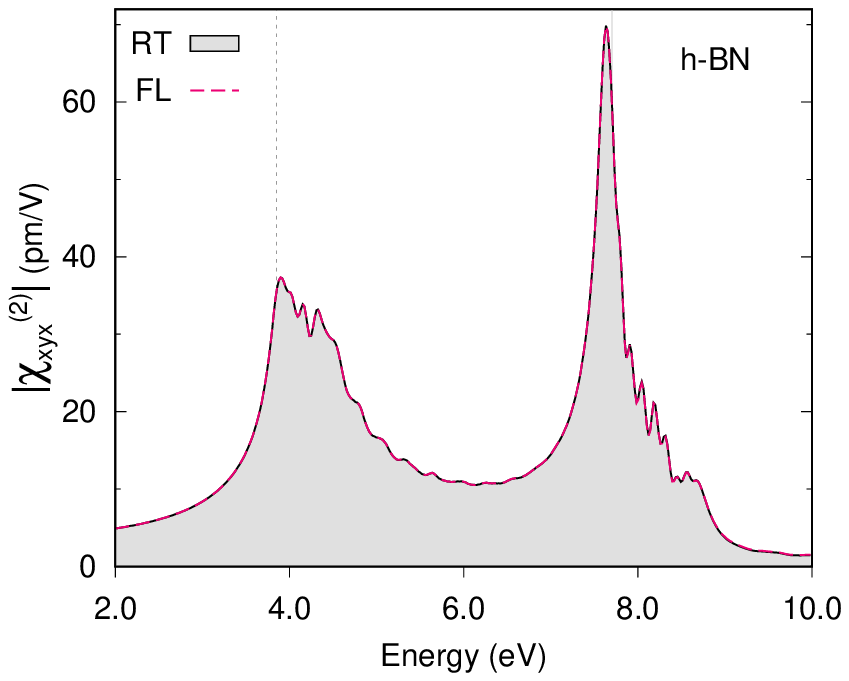}
    \caption{Monolayer h-BN SHG spectra on a $48\times48\times1$ $\kpt$-grid with a broadening of \SI{.15}{\eV}, calculated by the real-time (RT) approach (black line with grey filling) and our Floquet (FL) method (pink dashed line). The intensity of the electric field is 1$\times 10^3$ \si{\kilo\watt\per\centi\meter\squared}.}
    \label{fig:SHG_hBN}
\end{figure}

\begin{figure}[h!]
    \centering
    \includegraphics{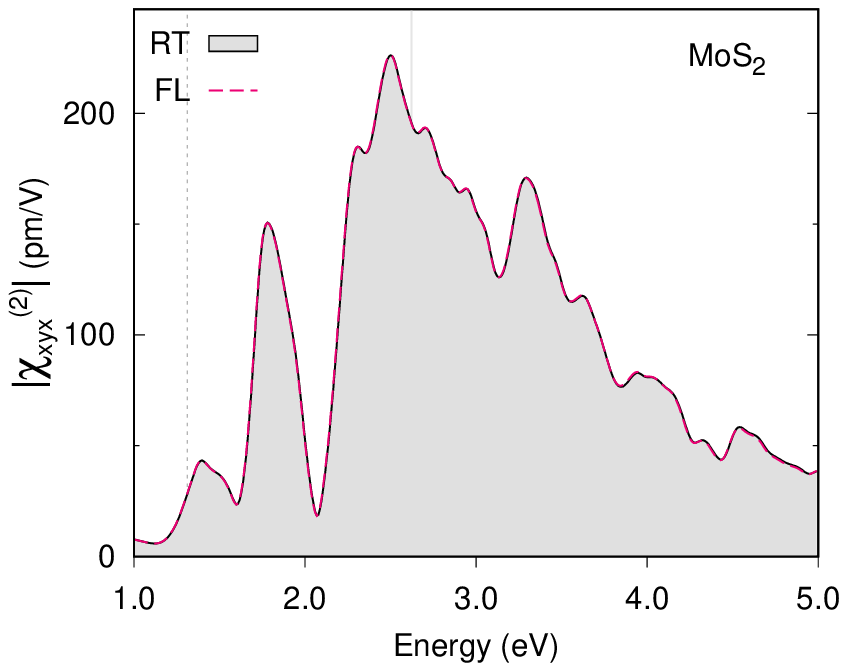}
    \caption{Monolayer MoS$_2$ SHG spectra on a $24\times24\times1$ $\kpt$-grid with a broadening of \SI{.15}{\eV}, calculated by the real-time (RT) approach (black line with grey filling) and our Floquet (FL) method (pink dashed line). The intensity of the electric field is 1$\times 10^3$ \si{\kilo\watt\per\centi\meter\squared}.}
    \label{fig:SHG_MoS2}
\end{figure}

In this section, we report selected SHG spectra for bulk AlAs, monolayer h-BN and monolayer MoS$_2$ calculated both \textit{via} the real-time and Floquet approaches (see computational details in Table \ref{tb:1}). Fig. \ref{fig:SHG_AlAs} presents SHG spectra for bulk AlAs calculated by both methods. The agreement between the two spectra is almost complete, despite the small broadening deliberately used to highlight differences. The small discrepancies towards the \SIrange{4}{6}{\eV} region are due to the choice of time-step in the real-time approach (see discussion below) and vanish with a shorter time step (see inset in Fig. \ref{fig:SHG_AlAs} and details in Table \ref{tb:1}). Fig. \ref{fig:SHG_hBN} shows the spectra for h-BN while Fig. \ref{fig:SHG_MoS2} presents data for MoS$_2$. It is apparent that the results produced by either method are indistinguishable from one another on this scale. This close matching between our Floquet method and the real-time approach was found across a variety of $\kpt$-point grids and broadening conditions for all three materials, extending also to the linear response regime (the full set of SHG and linear response results is provided in the \SuppInfo). 

\begin{table}
    \centering
    \begin{tabular}{l c c c }
        \hline
         & AlAs & h-BN & MoS$_2$ \\
        \hline
        $\kpt$-grids (nscf) & $\;30\times30\times30\;$ & $\;48\times48\times1\;$ & $\;24\times24\times1\;$\\
        Bands (full,empty) & 3 , 6 & 4 , 4 & 5 , 5\\
        Band-gap correction [\si{\eV}] & 0.9 & 3.3 & 0.72 \\
        Broadening [\si{\eV}] & 0.04 & 0.15 & 0.15 \\
        Total time [\si{\femto\second}] & 118 & 83 & 85 \\
        Time step [\si{\atto\second}]  & 10 (2.5) & 2.5 & 10 \\
        \hline
        
         \end{tabular}
    \caption{Computational details of the SHG calculations presented in Figs. \ref{fig:SHG_AlAs}-\ref{fig:SHG_MoS2}. The total time and time step have been selected through convergence tests available in the \SuppInfo. The time step in parentheses applies to the inset in Fig. \ref{fig:SHG_AlAs}}
    \label{tb:1}
\end{table}

Both real-time and Floquet calculations were carefully converged with respect to the relevant parameters in each case (see convergence tests in \SuppInfo). The Floquet approach requires convergence with respect to the number of Floquet modes included in each calculation, i.e., $\eta_\text{max}$ as defined in Eq. \ref{eq:eta_max_condition}. Our tests indicate that convergence with respect to $\eta_\text{max}$ is very fast for SHG spectra, e.g., $\eta_\text{max}=2$ is enough to compute a well-converged SHG spectrum, even at higher intensities where higher-order contributions should play a greater role (see convergence tests in \SuppInfo). In addition, our Floquet method requires an accuracy threshold to control the self-consistency loop. This threshold should be selected according to the magnitude of the response and will impact the computational cost of our approach through the number of iterations required. For SHG, we chose an accuracy threshold of $2.44\times10^{-5}$ \si{\pico\meter\per\volt}, which must be satisfied individually by both the real and imaginary parts of the second order susceptibility. This resulted in off-resonant frequencies converging within three iterations, while energies close to resonance took four or five iterations.

\begin{figure}[h!]
    \centering
    \includegraphics{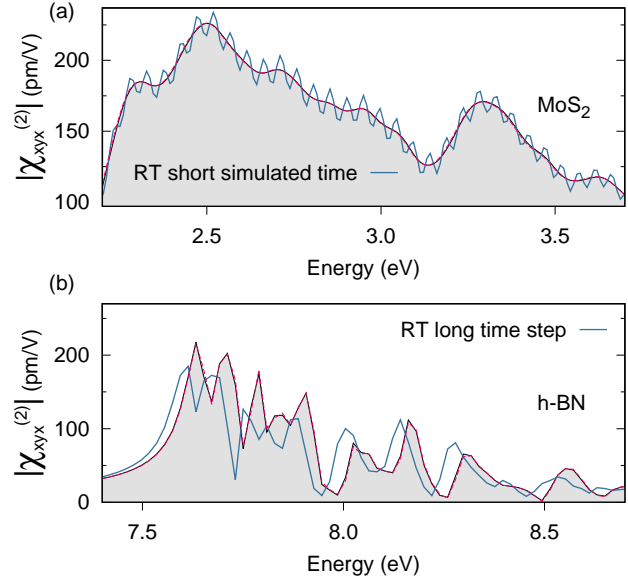}
    \caption{Effects of underconvergence with respect to total simulated time (a) and time step (b). Panel (a) shows the same spectra as Fig. \ref{fig:SHG_MoS2} plus an underconverged real-time (RT) spectrum obtained with \SI{65}{\femto\second} of total time (blue line). The latter compares poorly with the \SI{85}{\femto\second} required to properly dephase the system (black line with grey filling) and approach the Floquet solution (pink dashed line). Panel (b) displays the spectra for h-BN obtained with a well-converged time step of \SI{2.5}{\atto\second} (black line with grey filling), plus an underconverged real-time spectrum with \SI{10}{\atto\second} (blue line) that fails to approach the Floquet solution (pink dashed line). At variance with Fig. \ref{fig:SHG_hBN}, these spectra were calculated with a broadening of \SI{.04}{\eV} to highlight the differences.}
    \label{fig:RT-conv}
\end{figure}

Convergence in the case of real-time calculations must be studied with respect to two crucial parameters, the total simulated time and the time step used in the numerical integration of the EOMs. Both are system-dependent and must be subjected to convergence tests for every material. Regarding the former parameter, failure to allow for sufficient simulated time would result the response not being properly dephased. In this case, some eigenmodes of the system will still be excited since the introduction of the electric field at time zero, manifesting as oscillations in the real-time spectrum. For instance, the well converged spectrum in Fig. \ref{fig:SHG_MoS2} was obtained with \SI{85}{\femto\second} of total time. Fig. \ref{fig:RT-conv}(a) shows the same spectra of Fig. \ref{fig:SHG_MoS2} alongside an underconverged spectrum calculated with a total simulated time of \SI{65}{\femto\second}, which is evidently not enough to suppress these oscillations (see blue curve in Fig. \ref{fig:RT-conv}(a)). In terms of the time step, this needs to be sufficiently small since longer steps, albeit more efficient, will introduce unphysical features in the spectra. For instance, the well-converged h-BN spectrum in Fig. \ref{fig:SHG_hBN} was obtained with a time step of \SI{2.5}{\atto\second}. A portion of this spectrum was re-calculated with a low broadening (\SI{0.04}{\eV}) to highlight these unphysical features and is shown in Fig.  \ref{fig:RT-conv}(b). While the well-converged 2.5-\si{\atto\second} spectrum matches Floquet closely, the real-time spectrum integrated with a \SI{10}{\atto\second} step (thus, four times faster) is severely underconverged, as is apparent from the blue curve in Fig. \ref{fig:RT-conv}(b). Overall, real-time spectra tend to the Floquet solution upon progressively increasing the total simulated time and decreasing the time step, as inferred from Fig. \ref{fig:RT-conv}. Summarising, the convergence with numerical parameters impacts the execution time of the real-time approach to a larger extent with respect to the present formalism, which is not based on explicit time integration. The computational advantage of the present approach over the real-time one is demonstrated in Section \ref{sc:cost}).

\subsection{Third Harmonic Generation}\label{sc:THG}
\begin{figure}[h!]
    \centering
    \includegraphics{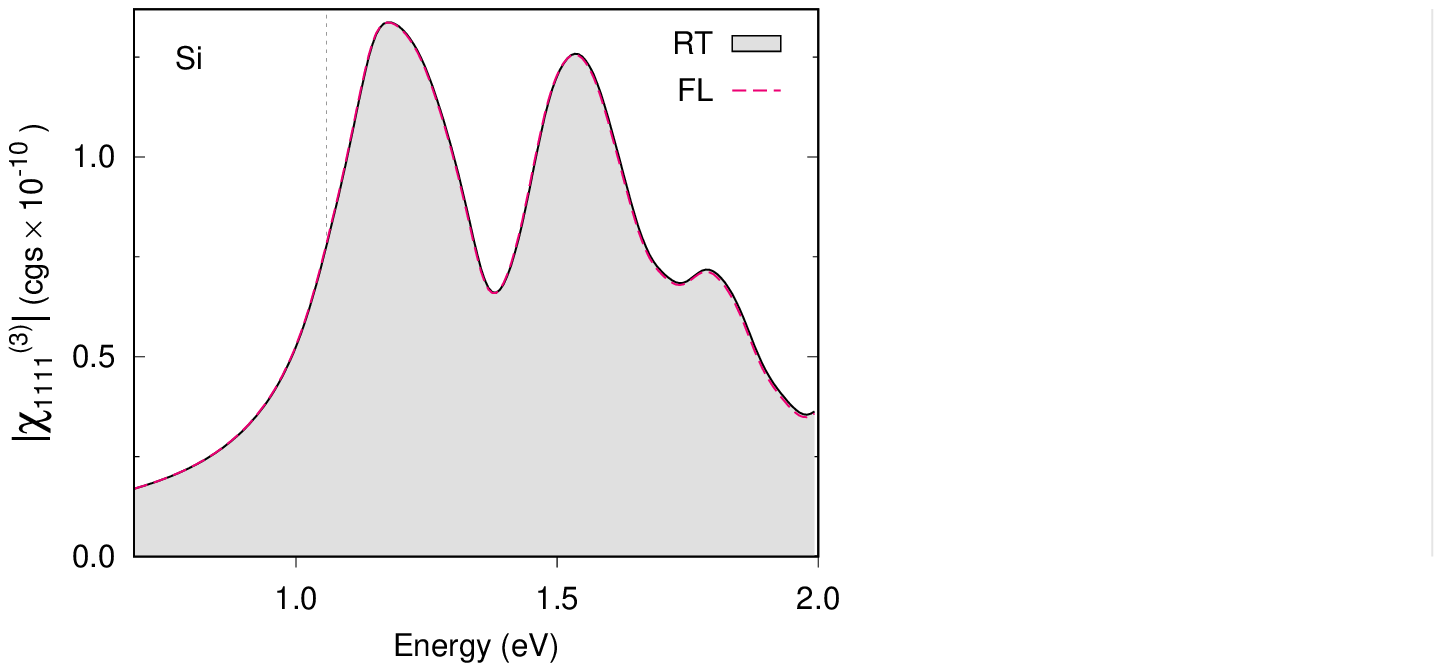}
    \caption{Bulk Si THG spectra on a $32\times32\times32$ $\kpt$-grid with a broadening of \SI{.15}{\eV}, calculated by the real-time (RT) approach (black line with grey filling) and our Floquet (FL) approach (pink dashed line). The intensity of the electric field is 1$\times 10^3$ \si{\kilo\watt\per\centi\meter\squared}. We used 4 occupied and 3 empty bands and a band gap correction of \SI{.6}{\eV}. The real-time spectrum was calculated with a total time of \SI{114}{\femto\second} and a time step of \SI{10}{\atto\second}. The Floquet spectrum was obtained with $\eta_\text{max}=3$.}
    \label{fig:THG}
\end{figure}

\begin{figure}[h!]
    \centering
    \includegraphics{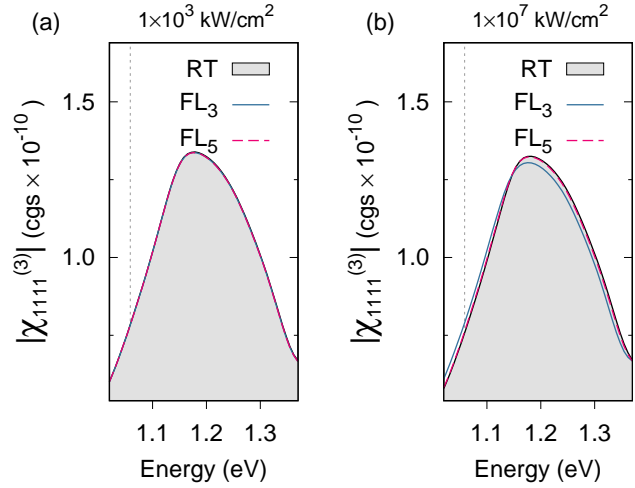}
    \caption{Higher-order contributions to THG spectra in bulk Si on a $32\times32\times32$ $\kpt$-grid. Spectra are calculated by both the real-time (RT) and Floquet (FL) approaches with electric fields of two different intensities, i.e., 1$\times 10^3$ (a) and 1$\times 10^7$ \si{\kilo\watt\per\centi\meter\squared} (b). The number of Floquet modes is indicated in the legend as a subscript, e.g., FL$_5$ corresponds to $\eta_\text{max}=5$. Computational details in Fig. \ref{fig:THG} apply, with two exceptions. A time step of \SI{1}{\atto\second} was used to rule out underconvergence in the real-time spectrum as a reason for the discrepancies. The total simulated times were set in excess of those required by the convergence tests for the same reason.}
    \label{fig:THG-5th-order}
\end{figure}

In this section, we report selected third harmonic generation (THG) spectra of bulk Si calculated both \textit{via} the real-time and Floquet approaches. The full set of THG results can be found in the \SuppInfo. Fig. \ref{fig:THG} shows very good agreement between the spectra calculated by either method. Convergence of the Floquet approach was also fast in this case, requiring only $\eta_\text{max}=3$ for THG spectra. However, additional Floquet modes may be required at higher intensities in case one wants to capture higher-order contributions to the third order response, which are present in the real-time result. These contributions depend on the intensity of the electric field and thus gain relevance at high intensities. In order to demonstrate this, we re-calculated a portion of these spectra with a perturbation of higher intensity and compared the results in Fig. \ref{fig:THG-5th-order}. The latter shows that $\eta_\text{max}=3$ suffices to reproduce the low-intensity real-time result while additional Floquet modes ($\eta_\text{max}=5$) are required at higher intensities to capture these higher-order contributions.  

\subsection{Computational cost}\label{sc:cost}
\begin{figure}[h!]
    \centering
    \includegraphics{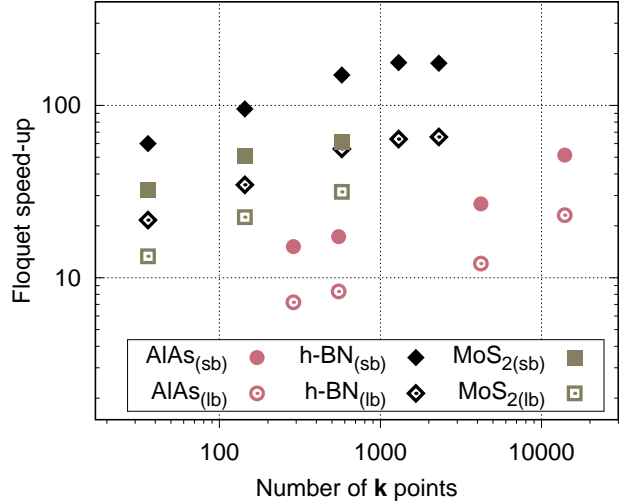}
    \caption{Computational cost comparison for SHG in the form of a CPU-time ratio between equivalent real-time and a Floquet calculations. Hollow markers represent calculations with \SI{.15}{\eV} broadening (``lb" stands for large broadening), while filled ones are for \SI{.04}{\eV} runs (``sb" or small-broadening). Floquet speed-ups are larger for the latter since the simulated times required are longer that in the former case, at the same time that Floquet CPU times are not significantly affected by the choice of broadening.}
    \label{fig:CompCost}
\end{figure}

We compared the computational cost of the real-time and Floquet approaches across the entire data set of SHG calculations. Our results account for a so-called Floquet speed-up of 1-2 orders of magnitude (see Fig. \ref{fig:CompCost}). This speed-up is calculated as the ratio of the CPU time required by either approach to perform the exact same calculation (see \SuppInfo \; for CPU time of each approach individually). Controlling the accuracy and convergence of the spectra played an important role in this comparison. As regards Floquet, we used $\eta_\text{max}=2$, which is well converged, and a uniform self-consistent accuracy threshold for all calculations. We then verified the latter was smaller than 0.1\% of the real-time result we intended to reproduce. We believe this is in line with what a regular user would do, however we note that there would have been potential for greater speed-ups had this threshold been fine-tuned in every calculation. As shown in Section \ref{sc:SHG}, the systematic way of converging real-time calculations implies increasing the total simulated time and decreasing the time step, i.e., two choices that increase the associated computational cost. While this was carefully tested for each material, we avoided overconverging these parameters as it would have unduly penalised the real-time approach (see \SuppInfo \; for computational details, convergence tests and all spectra in the data set). 

The results in Fig. \ref{fig:CompCost} show the influence of the real-time convergence parameters, e.g., the speed-up is higher in h-BN as it required the longest simulated times and shortest time steps (\SI{2.5}{\atto\second}) to closely match the Floquet spectra. MoS$_2$ shows an intermediate speed-up since it was well converged with a 10-\si{\atto\second} time step but also needed long simulated times. Finally, AlAs was calculated with a time step of \SI{10}{\atto\second}, which is well-converged in the region of interest despite the small discrepancies at \SIrange{4}{6}{\eV} (see Fig. \ref{fig:SHG_AlAs}). While reducing the latter with a \SI{2.5}{\atto\second} time step (see Fig. \ref{fig:SHG_AlAs}) would have quadrupled the Floquet speed-up, it would have unduly penalised the real-time approach in our view. 

One parameter that \textit{is} present in both methods and plays an important role in the speed-up achieved is the broadening of the spectra. In the real-time approach, the broadening is introduced through the dephasing term and impacts its ability to filter out excited eigenfrequencies. In general, a small broadening will require a long time to properly dephase the system and converge the spectrum. Hence, the broadening has an inverse impact in the computational cost of the real-time method through the total simulated time needed. At variance, the computational cost of our Floquet approach is almost insensitive to the broadening, which is introduced \textit{via} the damping term in Eq. \ref{eq:Damping}. It would be expected that a smaller broadening could make convergence more difficult at some particular (resonant) frequencies. However, while it is true that the CPU time required by Floquet scales linearly with the average number of self-consistent iterations per frequency, adding one or two cycles at a handful of frequencies did not impact the total CPU time significantly. As a result, the speed-up achieved with Floquet is much larger in small-broadening calculations. In fact, Fig. \ref{fig:CompCost} shows two groups of points per material. Within the data of a given material, the uppermost points correspond to small-broadening results (\SI{.04}{\eV}) while the lowermost ones reflect the large-broadening calculations (\SI{.15}{\eV}).

\begin{figure}[h!]
    \centering
    \includegraphics{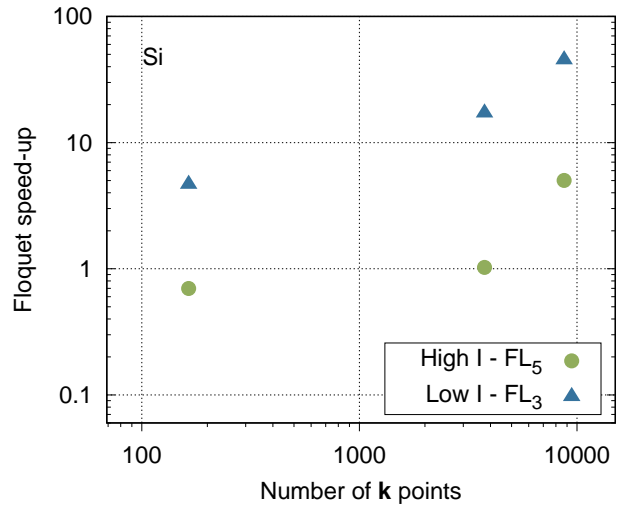}
    \caption{Computational cost comparison for bulk Si THG in the form of a CPU-time ratio between a real-time and a Floquet calculation producing the same spectra. In the labels, low and high I stand for intensities of $1\times10^3$ and $1\times10^7$ \si{\kilo\watt\per\centi\meter\squared}, respectively. The number of Floquet modes used is denoted by a subscript in the legend, e.g., FL$_5$ means $\eta_\text{max}=5$.}
    \label{fig:THGCompCost}
\end{figure}

We also performed comparisons on bulk Si THG spectra, which were calculated for just one broadening (\SI{.15}{\eV}) but two intensities, i.e., $1\times10^3$ and $1\times10^7$ \si{\kilo\watt\per\centi\meter\squared} (referred to as low and high intensity, respectively). In line with Section \ref{sc:THG}, we report high-intensity calculations with $\eta_\text{max}=5$ and low-intensity runs with $\eta_\text{max}=3$. This results in larger Floquet speed-ups at low intensities, as shown in Fig. \ref{fig:THGCompCost}. Since high-intensity spectra require $\eta_\text{max}=5$, there is an increased computational cost related to the additional Floquet modes in comparison with low-intensity calculations, for which $\eta_\text{max}=3$ is well converged (see \SuppInfo\; for convergence tests). Nonetheless, the computational advantage offered by Floquet becomes clear close to convergence with respect to $\kpt$-point sampling (see Fig. \ref{fig:THGCompCost}), regardless of the intensity. We point out that, with a broadening of \SI{.04}{\eV}, the low-intensity speed-up for an \kgrid{8}{8}{8} $\kpt$-grid was 15.1 (not shown in Fig. \ref{fig:THGCompCost}). This allows us to reliably estimate a speed-up of around 146 for a \kgrid{32}{32}{32} $\kpt$-grid. Hence, at low intensities where $\eta_\text{max}=3$ is well converged, the speed-ups obtained for bulk Si THG spectra are comparable to those achieved for bulk AlAs SHG spectra.

The drop in Floquet speed-ups at high intensities, i.e., when including higher harmonics to capture higher-order processes, points to a poor scaling of our method with respect to the number of Floquet modes. This is because each additional mode enlarges the dimension of the quasi-energy matrices by $2 N_b$, which in turns impacts the time required for their diagonalisation (see near-quadratic scaling in \SuppInfo). At the other end of the scale, linear response calculations are much more efficient with the present Floquet formalism than with the real-time approach. This is because the Floquet matrices are very small (of dimension $3 N_b$) and convergence very fast (typically 2 self-consistent iterations are sufficient for linear response). However, first-order Floquet is still more expensive than the usual frequency-domain response-based approach and the latter remains the best option in the linear response regime, at least at the independent particle level. 

The main contribution to the computational cost in our scheme is the diagonalisation of the quasi-energy matrices. For this particular task, we use the QR algorithm (i.e., `full' diagonalisation). There is room for improvement in this diagonalisation since the matrices are of dimension $N_b\times(2\eta_\text{max}+1)$ but only $M$ eigenvectors are needed (i.e., the number of occupied bands, which is a fraction of $N_b$). In the cases considered here, $M$ represents between 5 and $10 \%$ of the dimension of the corresponding matrices. This opens the possibility of exploring more efficient eigensolvers (e.g., those in the SLEPc library \cite{Hernandez:2005:SSF}) such as Krylov subspace methods \cite{Saad2011} or even variational approaches \cite{Kruger2020,Le2022}, which would further improve the performance of our Floquet method. Moreover, this would reduce the scaling of the computational cost with the number of Floquet modes.

\subsection{Limitations}
The limitations of the Floquet approach proposed in this manuscript are mainly related to the requirement of time-periodicity in the Hamiltonian. First and foremost, this framework can only apply to continuous and monochromatic perturbations. Otherwise, the basic conditions for the application of Floquet's theorem would not be present. For instance, modelling pump-and-probe experiments is beyond what one can do with the present formalism, and falls within the broader capabilities of the real-time approach. Second, we use the adiabatic approximation in order to ensure the periodicity of the effective Hamiltonian of our method, since the Berry-phase electron-field coupling operator depends self-consistently on the solution of the quasi-energy eigenproblem. While this limits the validity of the approach to weak field intensities, we performed tests up to $1\times10^7$ \si{\kilo\watt\per\centi\meter\squared} and found no signatures of non-adiabaticity. However, care should be taken when using our approach in this regime and beyond. In particular, we believe our method is not well-suited for the extreme nonlinear regime, where non-adiabaticity is expected to play an important role.

\section{Conclusions}\label{sc:conclusions}
In this work, we developed and implemented a Floquet approach to nonlinear optics for extended systems. This constitutes a reformulation of the real-time formalism \cite{Attaccalite2013} based on the Berry-phase dynamical polarisation \cite{Souza2004} and thus valid under PBCs. Exploiting the time-periodicity of the Hamiltonian in the presence of a monochromatic perturbation, we invoked Floquet's theorem to reformulate this time-dependent problem as a self-consistent time-independent eigenvalue problem. Our method applies to periodically-driven systems and is valid for weak electric fields, since we make use of the adiabatic approximation.

This Floquet formulation retains the non-perturbative nature of the real-time approach, allowing for the simultaneous extraction of susceptibilities to different orders in the electric field and for treating several nonlinear phenomena within the same formalism. Also, many-body effects can be included at different levels of approximation in our scheme by changing the effective one-particle Hamiltonian, e.g., excitonic effects can be included using a screened-exchange Coulomb-hole self energy or within density-polarisation functional theory, as it is done within the real-time approach \cite{Gruning2014,PhysRevB.94.035149}. On the other hand, our reformulation tackles the often prohibitive computational cost associated with real-time calculations, which originates from the expensive numerical integration of the EOMs (often requiring very short time steps). This cumbersome time propagation is entirely avoided in our time-independent formalism.

We demonstrated the validity and effectiveness of the proposed Floquet scheme by implementing it at the independent particle level and testing it extensively on a number of well-studied materials. We calculated optical absorption and SHG spectra of bulk AlAs, monolayer h-BN and monolayer MoS$_2$, plus THG spectra of bulk Si. In all cases, the Floquet method produced spectra in agreement with the implementation of the real-time approach in Ref.~\cite{Attaccalite2013,Sangalli2019}. The proposed scheme showed a consistent computational advantage in comparison to the real-time formalism, resulting up to two orders of magnitude faster. A further computational speed-up could be achieved by employing a more efficient eigensolver for the quasi-energy eigenproblem, as those available from the SLEPc library~\cite{Hernandez:2005:SSF}. In light of these results, our contribution holds the promise to enable the \textit{ab-initio} calculation of nonlinear optical properties for a range of complex materials that are too demanding for currently available methods.

\section*{Author Contributions}
MG put forward the idea for the approach, developed it and assisted with its implementation into Yambo. MG also contributed to the analysis of the results and reviewed the manuscript.
IMA contributed to the development of the approach and implemented it into Yambo. IMA also carried out the calculations, analysed the results and wrote the manuscript. 

\section*{Acknowledgments}
The authors acknowledge useful discussions with Nicolas Tancogne-Dejean and Daniel Dundas. The authors are grateful for use of the computing resources from the Northern Ireland High Performance Computing (NI-HPC) service funded by the Engineering and Physical Sciences Research Council (EP/T022175) and from the School of Mathematics and Physics at Queen's University Belfast.
MG is grateful for support from the Engineering and Physical Sciences Research Council, under grant EP/V029908/1.
The authors declare that the research was conducted in the absence of any commercial or financial relationships that could be construed as a potential conflict of interest.

\section*{References}
\bibliographystyle{iopart-num}
\bibliography{library.bib,books.bib}

\end{document}